# Dilaton and Second-Rank Tensor Fields as Supersymmetric Compensators

Hitoshi Nishino[1)] and Subhash Rajpoot[2)]

*Department of Physics & Astronomy*
*California State University*
*1250 Bellflower Boulevard*
*Long Beach, CA 90840*

## Abstract

We formulate a supersymmetric theory in which both a dilaton and a second-rank tensor play roles of compensators. The basic off-shell multiplets are a linear multiplet $(B_{\mu\nu}, \chi, \varphi)$ and a vector multiplet $(A_\mu, \lambda; C_{\mu\nu\rho})$, where $\varphi$ and $B_{\mu\nu}$ are respectively a dilaton and a second-rank tensor. The third-rank tensor $C_{\mu\nu\rho}$ in the vector multiplet is 'dual' to the conventional $D$-field with 0 on-shell or 1 off-shell degree of freedom. The dilaton $\varphi$ is absorbed into one longitudinal component of $A_\mu$, making it massive. Initially, $B_{\mu\nu}$ has 1 on-shell or 3 off-shell degrees of freedom, but it is absorbed into the longitudinal components of $C_{\mu\nu\rho}$. Eventually, $C_{\mu\nu\rho}$ with 0 on-shell or 1 off-shell degree of freedom acquires in total 1 on-shell or 4 off-shell degrees of freedom, turning into a propagating massive field. These basic multiplets are also coupled to chiral multiplets and a supersymmetric Dirac-Born-Infeld action. Some of these results are also reformulated in superspace. The proposed mechanism may well provide a solution to the long-standing puzzle of massless dilatons and second-rank tensors in supersymmetric models inspired by string theory.

PACS: 11.15.-q, 11.30.Pb, 12.60.Jv

Key Words: Supersymmetry, Stueckelberg and Proca Formalisms, Compensators, Dilaton, Second-Rank Tensor, Third-Rank Tensor, Dirac-Born-Infeld Action.

[1)] E-Mail: hnishino@csulb.edu
[2)] E-Mail: rajpoot@csulb.edu

## 1. Introduction

In higher dimensional theories, such as superstring theory [1] or supergravity theory [2][3], compactifications of extra dimensions generally result in many massless fields known as moduli fields. They couple with gravitational strength and determine the matter couplings in the theory. However, such massless particles are inconsistent with experiment, as they have not been observed in nature. In spite of generating effective potentials for the particles that lend to physical interpretations, these massless fields cannot be easily fixed.

For example, the dilaton field arises in superstring theory [1] as the massless scalar field in the Neveu-Schwarz (NS) sector. In supergravity theory [2] as its low-energy limit, the dilaton inevitably arises as exponential factors associated with global scaling symmetry in matter fields. From a Kaluza-Klein theory viewpoint, the dilaton arises as an exponential factor in the direction of extra dimensions. From these considerations, the dilaton is natural and indispensable in any of these important theories. However, the massless dilaton is incompatible with cosmological observations [1].

Another problematic field is an antisymmetric tensor $B_{\mu\nu}$ in the NS sector generating the second-rank (two-form) field with properties similar to the axion, if one seeks to solve the strong CP problem in QCD *via* the Peccei-Quinn mechanism [4][5]. The decay constant $f$ of such axions is expected to be around the string scale: $10^{16}$ GeV $\lesssim f \lesssim 10^{19}$ GeV. This poses a serious problem, as these values lie outside the allowed range on axion couplings. Astrophysical data suggest $f \gtrsim 10^9$ GeV [6]. This implies that the axion-like field must be very light and extremely weakly coupled. On the other hand, cosmological arguments on the overclosure of the universe also yield an upper bound $f \lesssim 10^{12}$ GeV [7]. Of course, these constraints are pertinent, provided the right conditions are implemented in the model for these axion-like particles to address the strong CP problem. However, there are ways to circumvent the shortcomings posed by these particles [5].

In this paper, we propose a new model in which both a dilaton and a two-form field play roles of compensators, being absorbed into certain tensors, and disappear from physical particle spectrum. We further supersymmetrize this mechanism, such that the dilaton and two-form fields play roles of compensators, and are absorbed into certain tensors. This mechanism is similar to a compensator, Proca or Stueckelberg formalism [8][9][10][11]. The formulation we present in this paper is also similar to our previous results on supersymmetric compensator in 3D [12]. Namely, we show that the dilaton $\varphi$ can play a role of a compensator for a vector field $A_\mu$ in 4D. In other words, the 0-th rank tensor $\varphi$ can be absorbed into a 1-st rank tensor $A_\mu$. Analogously, we show that the two-form field $B_{\mu\nu}$ can also play a role of a compensator for a three-form tensor $C_{\mu\nu\rho}$ simultaneously.

Note that the technique of compensator fields itself is nothing new ever since the original works by Proca [9] and Stueckelberg [8][10][11]. For example, in the so-called anti-Higgs mechanism [13] a massless field is eaten up by an antisymmetric tensor which thereby describes a massive spin 1 field. Another well-known example is massive type IIA supergravity [14] in which a vector field is absorbed into a second-rank tensor which becomes massive. However, in the model we present here, not only the dilaton $\varphi$, but also the second-rank antisymmetric tensor $B_{\mu\nu}$ will be absorbed respectively into a vector $A_\mu$ and a third-rank tensor $C_{\mu\nu\rho}$ at the same time.

The supermultiplets we consider are the linear multiplet (LM) $(B_{\mu\nu}, \chi, \varphi)$ [15], and the vector multiplet (VM) $(A_\mu, \lambda; C_{\mu\nu\rho})$. In the LM, $\varphi$ and $B_{\mu\nu}$ are regarded as a dilaton and a two-form field, respectively. The former is absorbed into a vector $A_\mu$, while the latter is absorbed into a three-form tensor $C_{\mu\nu\rho}$, which is originally an auxiliary field in the VM $(A_\mu, \lambda; C_{\mu\nu\rho})$. The conventional VM $(A_\mu, \lambda; D)$ has an auxiliary field $D$. However, we 'dualize' $D$ into $C_{\mu\nu\rho}$, which will be massive and physically propagating.

As has been well known, the original $C_{\mu\nu\rho}$ field has zero on-shell degree of freedom in 4D, and therefore it is *not* propagating. In fact, the on-shell counting works as $(4-2)(4-3)(4-4)/3! = 0$, while the off-shell counting is $(4-1)(4-2)(4-3)/3! = 1$. However, in our mechanism, the two-form field $B_{\mu\nu}$ plays a role of compensator field, and is absorbed into the longitudinal components of $C_{\mu\nu\rho}$, making it massive. For $B_{\mu\nu}$, the on-shell counting is $(4-2)(4-3)/2! = 1$, while the off-shell counting is $(4-1)(4-2)/2! = 3$. Therefore the latter 1 on-shell or 3 off-shell degrees of freedom are absorbed into $C_{\mu\nu\rho}$, resulting in 1 on-shell or 4 off-shell degrees of freedom. These results are recapitulated in the following table:

| Degrees of Freedom | $B_{\mu\nu}$ | $C_{\mu\nu\rho}$ | Massive $C_{\mu\nu\rho}$ |
|:---:|:---:|:---:|:---:|
| On-Shell | 1 | 0 | 1 |
| Off-Shell | 3 | 1 | 4 |

Table 1: Degrees of Freedom for $B_{\mu\nu}$ and $C_{\mu\nu\rho}$

The important point here is that the conventionally frozen field $C_{\mu\nu\rho}$ becomes massive and propagating, after the absorption of the two-form field $B_{\mu\nu}$ like a compensator field [8][10][11].

As the first non-trivial interactions, we couple our basic system to chiral multiplets, and next to supersymmetric Dirac-Born-Infled (SDBI) action [16]. These non-trivial interactions imply that our system is a significant physical system, in particular, the massive three-form tensor $C_{\mu\nu\rho}$ is combined with the second-rank tensor $B_{\mu\nu}$ consistently with supersymmetry. The possibility of spontaneous supersymmetry breaking is also studied. Subsequently, we reformulate the component result in superspace. Interestingly, we will see that an 'auxiliary' superfield strength $L_{ABC}$ plays a crucial role for the satisfaction of Bianchi identities.

## 2. Preliminaries about Three-Form Field $C_{\mu\nu\rho}$

We first understand the mechanism for the massive propagating tensor $C_{\mu\nu\rho}$, when the two-form field $B_{\mu\nu}$ is absorbed into the longitudinal components of $C_{\mu\nu\rho}$. To this end, we consider the toy action $I_H \equiv \int d^4x\, \mathcal{L}_H$ with[3]

$$\mathcal{L}_H = -\tfrac{1}{48}(H_{\mu\nu\rho\sigma})^2 - \tfrac{1}{12}(G_{\mu\nu\rho})^2 \ . \tag{2.1}$$

The field strengths $H$ and $G$ are defined by

$$H_{\mu\nu\rho\sigma} \equiv 4\partial_{[\mu} C_{\nu\rho\sigma]} \ , \tag{2.2a}$$

$$G_{\mu\nu\rho} \equiv 3\partial_{[\mu} B_{\nu\rho]} + mC_{\mu\nu\rho} \equiv +m\widetilde{C}_{\mu\nu\rho} \ . \tag{2.2b}$$

The latter shows that the $B$-field can be absorbed into the field redefinition $\widetilde{C}_{\mu\nu\rho} \equiv C_{\mu\nu\rho} + 3m^{-1}\partial_{[\mu} B_{\nu\rho]}$. In other words, $B_{\mu\nu}$ plays a role of a compensator field. Now, the lagrangian $\mathcal{L}_H$ is equivalent to that of a Proca-type [9] *massive* field $\widetilde{C}_{\mu\nu\rho}$:

$$\mathcal{L}_H = -\tfrac{1}{48}(\widetilde{H}_{\mu\nu\rho\sigma})^2 - \tfrac{1}{12}m^2(\widetilde{C}_{\mu\nu\rho})^2 \ , \tag{2.3}$$

yielding the field equation[4]

$$\partial_\sigma \widetilde{H}_{\mu\nu\rho}{}^\sigma + m^2 \widetilde{C}_{\mu\nu\rho} \doteq 0 \ . \tag{2.4}$$

---

[3] In this paper, we use the metric $(\eta_{\mu\nu}) = \text{diag.}(-,+,+,+)$.
[4] We use the symbol $\doteq$ for a field equation in this paper.

In order to simplify the analysis, we introduce the 'dual' field $v_\mu$ defined by

$$v_\mu \equiv +\tfrac{1}{6}\epsilon_\mu{}^{\rho\sigma\tau}\widetilde{C}_{\rho\sigma\tau} \quad, \qquad \widetilde{C}_{\mu\nu\rho} = +\epsilon_{\mu\nu\rho}{}^\sigma v_\sigma \quad, \tag{2.5}$$

so that $\widetilde{H}$ is expressed as $\widetilde{H}_{\mu\nu\rho\sigma} = -4\epsilon_{[\mu\nu\rho}{}^\tau \partial_{\sigma]} v_\tau$. Eventually, $\mathcal{L}_H$ becomes

$$\mathcal{L}'_H = +\tfrac{1}{2}(\partial_\mu v^\mu)^2 + \tfrac{1}{2}m^2 v_\mu^2 \quad. \tag{2.6}$$

Now the $v$-field equation from (2.6) is consistent with the one obtained by substituting (2.5) into (2.4), as

$$\partial_\mu \partial_\nu v^\nu - m^2 v_\mu \doteq 0 \quad. \tag{2.7}$$

If $m \neq 0$, we can solve (2.7) for $v$ as

$$v_\mu \doteq + m^{-2} \partial_\mu \partial_\nu v^\nu = +m^{-2}\partial_\mu \phi \quad. \tag{2.8}$$

where $\phi$ is a scalar field defined by

$$\phi \equiv \partial_\mu v^\mu \quad, \tag{2.9}$$

Eq. (2.8) implies that $v_\mu$ is nothing but a gradient $v_\mu \doteq + m^{-2}\partial_\mu\phi$.

Interestingly, when (2.8) is substituted back into the original $\widetilde{C}$-field equation (2.4), it yields

$$\partial_\mu(\partial_\nu^2 \phi - m^2 \phi) \doteq 0 \quad. \tag{2.10}$$

Since the index $\mu$ is free, (2.10) implies nothing but the massive Klein-Gordon equation for the scalar $\phi$, under the usual boundary condition $\phi \to 0$ at spacial infinities $|x^i| \to \infty$. In other words, after absorbing the two-form field $B_{\mu\nu}$, the original field $C_{\mu\nu\rho}$ becomes massive and propagating in 4D. Note the non-trivial fact that the mass term in (2.10) has the right sign instead of a tachyonic mass. This is other supporting evidence that our formulation is the right one for the massive three-form tensor $C_{\mu\nu\rho}$.

To our knowledge, this feature has not been pointed out in the past, in the context of neither two-form nor three-form tensors. The common wisdom tells us that a three-form tensor $C_{\mu\nu\rho}$ is not an interesting field in 4D, because it has no on-shell degree of freedom, and is not propagating anyway. The only interest in the three-form tensor in 4D has been for compactifications of 11D supergravity [17]. However, we have found above that this seemingly-frozen field is revived and starts propagating, after absorbing the whole field of the two-form tensor $B_{\mu\nu}$ as its longitudinal components.

## 3. Supersymmetric Compensators $\varphi$ and $B_{\mu\nu}$

As has been already mentioned, we have the two fundamental multiplets: a LM $(B_{\mu\nu}, \chi, \varphi)$ [15] and a VM $(A_\mu, \lambda; C_{\mu\nu\rho})$. Note that both of them are off-shell multiplets, with $2+2$ on-shell or $4+4$ off-shell degrees of freedom.

Our basic action $I_0 \equiv \int d^4 x\, \mathcal{L}_0$ has the lagrangian

$$\begin{aligned}\mathcal{L}_0 = &-\tfrac{1}{48}(H_{\mu\nu\rho\sigma})^2 - \tfrac{1}{12}(G_{\mu\nu\rho})^2 - \tfrac{1}{4}(F_{\mu\nu})^2 - \tfrac{1}{2}(D_\mu\varphi)^2 \\ &+ \tfrac{1}{2}(\overline{\lambda}\slashed{\partial}\lambda) + \tfrac{1}{2}(\overline{\chi}\slashed{\partial}\chi) + m(\overline{\lambda}\chi) ~,\end{aligned} \quad (3.1)$$

where the covariant derivative $D\varphi$ and the field strengths $F$, $G$ and $H$ are defined by

$$D_\mu\varphi \equiv +\partial_\mu\varphi + mA_\mu ~, \quad (3.2a)$$

$$F_{\mu\nu} \equiv +2\partial_{[\mu} A_{\nu]} ~, \quad (3.2b)$$

$$G_{\mu\nu\rho} \equiv +3\partial_{[\mu} B_{\nu\rho]} + mC_{\mu\nu\rho} ~, \quad (3.2c)$$

$$H_{\mu\nu\rho\sigma} \equiv +4\partial_{[\mu} C_{\nu\rho\sigma]} ~, \quad (3.2d)$$

where (3.2c) is the same as (2.2b). The field strengths $D\varphi$ and $G$ satisfy the non-trivial Bianchi identities

$$\partial_{[\mu} D_{\nu]}\varphi \equiv +\tfrac{1}{2} m F_{\mu\nu} ~, \quad (3.3a)$$

$$\partial_{[\mu} G_{\nu\rho\sigma]} \equiv +\tfrac{1}{4} m H_{\mu\nu\rho\sigma} ~. \quad (3.3b)$$

Our action $I_0$ is invariant under supersymmetry

$$\delta_Q B_{\mu\nu} = +(\overline{\epsilon}\gamma_{\mu\nu}\chi) ~, \quad (3.4a)$$

$$\delta_Q \chi = -(\gamma^\mu\epsilon) D_\mu\varphi + \tfrac{1}{6}(\gamma^{\mu\nu\rho}\epsilon) G_{\mu\nu\rho} ~, \quad (3.4b)$$

$$\delta_Q \varphi = +(\overline{\epsilon}\chi) ~, \quad (3.4c)$$

$$\delta_Q A_\mu = +(\overline{\epsilon}\gamma_\mu\lambda) ~, \quad (3.4d)$$

$$\delta_Q \lambda = +\tfrac{1}{2}(\gamma^{\mu\nu}\epsilon) F_{\mu\nu} - \tfrac{1}{24}(\gamma^{\mu\nu\rho\sigma}\epsilon) H_{\mu\nu\rho\sigma} ~, \quad (3.4e)$$

$$\delta_Q C_{\mu\nu\rho} = +(\overline{\epsilon}\gamma_{\mu\nu\rho}\lambda) ~. \quad (3.4f)$$

Since we are using the off-shell formulation, supersymmetry closes without field equations.

Our action $I_0$ is also invariant under the local infinitesimal gauge symmetries

$$\delta_\xi \varphi = -m\xi \ , \tag{3.5a}$$

$$\delta_\xi A_\mu = +\partial_\mu \xi \ , \tag{3.5b}$$

$$\delta_{\eta,\zeta} B_{\mu\nu} = +2\partial_{[\mu}\eta_{\nu]} - m\zeta_{\mu\nu} \ , \tag{3.5c}$$

$$\delta_\zeta C_{\mu\nu\rho} = +3\partial_{[\mu}\zeta_{\nu\rho]} \ . \tag{3.5d}$$

It is by the property (3.5a) that we can call $\varphi$ 'dilaton', corresponding to the constant shift $\varphi = -m\xi$ in the global case. By the same token, under the $\eta$-transformation (3.5c), $B_{\mu\nu}$ shares the same property with an 'axion' [4].

## 4. Coupling to Chiral Multiplets

Since we have established a free supersymmetric system, the next natural step is to consider certain interactions. The simplest example is the coupling to a pair of chiral multiplets forming the **2** of $SO(2)$: $(A^i, B^i, \chi^i; F^i, G^i)$, where $i, j, \cdots = 1, 2$ are for the **2** of $SO(2)$. Their supersymmetry transformation rule is

$$\delta_Q A^i = +(\overline{\epsilon}\chi^i) \ , \quad \delta_Q B^i = +i(\overline{\epsilon}\gamma_5\chi^i) \ , \tag{4.1a}$$

$$\delta_Q \chi^i = -(\gamma^\mu \epsilon) D_\mu A^i + i(\gamma_5 \epsilon) D_\mu B^i - \epsilon F^i - i(\gamma_5 \epsilon) G^i \ , \tag{4.1b}$$

$$\delta_Q F^i = +(\overline{\epsilon}\slashed{D}\chi^i) + g\epsilon^{ij}(\overline{\epsilon}\lambda)A^j - ig\epsilon^{ij}(\overline{\epsilon}\gamma_5\lambda)B^j \ , \tag{4.1c}$$

$$\delta_Q G^i = +i(\overline{\epsilon}\gamma_5 \slashed{D}\chi^i) - g\epsilon^{ij}(\overline{\epsilon}\lambda)B^j + ig\epsilon^{ij}(\overline{\epsilon}\gamma_5\lambda)A^j \ . \tag{4.1c}$$

As usual, the $SO(2)$-covariant derivative $D_\mu$ is defined by

$$D_\mu A^i \equiv \partial_\mu A^i + g\epsilon^{ij} A_\mu A^j \ , \quad D_\mu B^i \equiv \partial_\mu B^i + g\epsilon^{ij} A_\mu B^j \ , \tag{4.2a}$$

$$D_\mu \chi^i \equiv \partial_\mu \chi^i + g\epsilon^{ij} A_\mu \chi^j \ . \tag{4.2b}$$

An invariant action $I_{\rm CM} \equiv \int d^4x\, \mathcal{L}_{\rm CM}$ for the kinetic terms has the lagrangian

$$\begin{aligned}\mathcal{L}_{\rm CM} = & -\tfrac{1}{2}(D_\mu A^i)^2 - \tfrac{1}{2}(D_\mu B^i)^2 + \tfrac{1}{2}(\overline{\chi}^i \slashed{D}\chi^i) + \tfrac{1}{2}(F^i)^2 + \tfrac{1}{2}(G^i)^2 \\ & + g\epsilon^{ij}(\overline{\lambda}\chi^i)A^j + ig\epsilon^{ij}(\overline{\lambda}\gamma_5\chi^i)B^j - g\epsilon^{ij} H A^i B^j \ ,\end{aligned} \tag{4.3}$$

where the pseudoscalar field $H$ is dual to $H_{\mu\nu\rho\sigma}$:

$$H \equiv +\tfrac{1}{24}\epsilon^{\mu\nu\rho\sigma} H_{\mu\nu\rho\sigma} \quad . \tag{4.4}$$

The invariance $\delta_Q I_{\rm CM} = 0$ is not too difficult to confirm. In particular, the last term with $H$ in (4.3) contributes to the three sectors: (i) $gAH\chi$, (ii) $gBH\chi$, (iii) $g\lambda ADB$ and (iv) $g\lambda BDA$. These are all cancelled by the like terms generated by the $g\lambda\chi A$ and $g\lambda\chi B$-terms in the lagrangian.

Basically, the interaction structure in this lagrangian is parallel to the conventional case with all the $D$-field is replaced by the $H$-field. Despite of this parallel structure, we stress also the important difference due to the field strength $H_{\mu\nu\rho\sigma}$ involved in all the $H$-dependent terms. Since we are adopting an off-shell formulation, the lagrangian $\mathcal{L}_{\rm CM}$ can be added to $\mathcal{L}_0$ in (3.1) without disturbing the invariance of the total action $I_1 \equiv I_0 + I_{\rm CM}$.

## 5.  A Test of Spontaneous Supersymmetry Breaking

In a conventional supersymmetry theory, we can add the so-called Fayet-Ilyapoulos term $\mathcal{L}_{\rm FI} \equiv \xi D$ [18] with the VM auxiliary field $D$ to an arbitrary supersymmetric lagrangian. Then the $D$-field equation will be $D \doteq -\xi$ breaking supersymmetry spontaneously, because the $D$ enters in the variation $\delta_Q \lambda$ as $i\gamma_5 \epsilon D$, signaling that $\lambda$ is a Nambu-Goldstino. We can perform a similar analysis for our system. For example, we can add an analogous term

$$\mathcal{L}_{\xi H} = \tfrac{1}{24}\xi\,\epsilon^{\mu\nu\rho\sigma} H_{\mu\nu\rho\sigma} = \xi H \tag{5.1}$$

to our $\mathcal{L}_0$. However, $\mathcal{L}_{\xi H}$ is a total divergence, affecting no field equation.

We first get the $C$ and $B$-field equations from the total action $I_2 \equiv I_0 + I_{\xi H}$, as

$$\partial_\sigma H^{\mu\nu\rho\sigma} + m G^{\mu\nu\rho} \doteq 0 \quad , \tag{5.2a}$$

$$\partial_\rho G^{\mu\nu\rho} \doteq 0 \quad . \tag{5.2b}$$

The former allows us to solve it for $G$ as

$$G_{\mu\nu\rho} \doteq -\,m^{-1}\partial_\sigma H_{\mu\nu\rho}{}^\sigma = +m^{-1}\epsilon_{\mu\nu\rho}{}^\sigma \partial_\sigma H \quad . \tag{5.3}$$

We next look into the dynamical energy-momentum tensor for $G$ and $H$:

$$T_{\mu\nu}\Big|_{G,H} \equiv -2e^{-1}\frac{\delta\mathcal{L}_2}{\delta g^{\mu\nu}}\bigg|_{G,H} = -\tfrac{1}{2}\eta_{\mu\nu}H^2 - \tfrac{1}{6}\eta_{\mu\nu}G^2_{\rho\sigma\tau} + \tfrac{1}{2}G_\mu{}^{\rho\sigma}G_{\nu\rho\sigma} ~, \tag{5.4}$$

where we temporarily introduced the metric $g_{\mu\nu}$ to the kinetic terms, and after the variation, we went back to the flat metric $\eta_{\mu\nu}$. Due to the absence of the metric in $\epsilon^{\mu\nu\rho\sigma}$, $\mathcal{L}_{\xi H}$ does *not* contribute to $T^{\mu\nu}$. We now substitute (5.3) into (5.4), eliminating $G$, as

$$T_{\mu\nu}\Big|_{G,H} \doteq -\tfrac{1}{2}\eta_{\mu\nu}H^2 - \tfrac{1}{2}m^{-2}\eta_{\mu\nu}(\partial_\rho H)^2 + m^{-2}(\partial_\mu H)(\partial_\nu H) ~. \tag{5.5}$$

As desired, the 00-component of $T^{00}$ is positive definite:

$$T^{00}\Big|_{G,H} \doteq +\tfrac{1}{2}H^2 + \tfrac{1}{2}m^{-2}(\partial_i H)^2 + \tfrac{1}{2}m^{-2}\dot{H}^2 \geq 0 ~, \tag{5.6}$$

with the spacial coordinate index $i = 1, 2, 3$. This $T^{00}$ is minimized to zero, only if

$$H \doteq 0 ~, \tag{5.7}$$

Since $H$ is involved in the supersymmetry transformation $\delta_Q \lambda$ (3.4e) as $i\gamma_5 \epsilon H$, supersymmetry is intact for the solution $H \doteq 0$.

In general spontaneous symmetry breaking, whereas the lagrangian itself or field equations are invariant under a given symmetry, a solution giving the minimal value of energy breaks the symmetry. In our case, the situation is as follows. Among all the possible solutions of $H$, only $H \doteq 0$ minimizes energy, maintaining supersymmetry as the vacuum solution. In terms of initial and boundary conditions on $H(x^i, t)$ at $|x^i| \to \infty$ and $t \to -\infty$, only $H(\pm\infty, -\infty) = 0$ minimizes $T^{00}$, maintaining supersymmetry. This is also consistent with the fact that our Fayet-Iliopoulos-like term $\mathcal{L}_{\xi H}$ is a total divergence with *no* effect of supersymmetry breaking.

Even though $\mathcal{L}_{\xi H}$ itself does not break supersymmetry spontaneously, the usual O'Raifearteigh mechanism [19] with additional chiral multiplets works just fine in our model as well.

## 6. Coupling to SDBI Action

As another example of non-trivial interactions, we show the couplings to SDBI action [16]. Our VM $(A_\mu, \lambda; C_{\mu\nu\rho})$ with $C_{\mu\nu\rho}$ instead of $D$ reveals a slight difference. However, as far as the SDBI action [16] is concerned, such a difference will not pose any problem, because the dual $H \equiv (1/4!)\epsilon^{\mu\nu\rho\sigma} H_{\mu\nu\rho\sigma}$ of $H_{\mu\nu\rho\sigma}$ replaces all the $D$-field involved in the usual SDBI action [16], while keeping the total action invariant.

After these arrangements, we get the SDBI action $I_{\text{SDBI}} \equiv \int d^4x\, \mathcal{L}_{\text{SDBI}}$, given by

$$\begin{aligned}
\mathcal{L}_{\text{SDBI}} =\ & +\tfrac{1}{4}\alpha'(F^4)_\mu{}^\mu - \tfrac{1}{16}\alpha'(F^2_{\mu\nu})^2 + \tfrac{1}{4}\alpha' H^4 - \tfrac{1}{4}\alpha' F^2_{\mu\nu} H^2 \\
& - \alpha'(\overline{\lambda}_+ \partial_\mu \lambda_+)(\overline{\lambda}_- \partial^\mu \lambda_-) + \alpha'(\overline{\lambda}_+ \partial\!\!\!/ \lambda_-)(\overline{\lambda}_- \partial\!\!\!/ \lambda_+) \\
& + \tfrac{1}{2}\alpha'\bigl[ (\overline{\lambda}_+ \partial\!\!\!/ \lambda_-) + (\overline{\lambda}_- \partial\!\!\!/ \lambda_+) \bigr]\bigl(H^2 - \tfrac{1}{2}F^2_{\mu\nu}\bigr) + \tfrac{i}{4}\alpha'\bigl[ (\overline{\lambda}_+ \partial\!\!\!/ \lambda_-) - (\overline{\lambda}_- \partial\!\!\!/ \lambda_+) \bigr] F_{\mu\nu}\widetilde{F}^{\mu\nu} \\
& + \tfrac{1}{4}\alpha'\bigl[ \overline{\lambda}_+ H + \tfrac{i}{2}(\overline{\lambda}_+ \gamma^{\mu\nu}) F_{\mu\nu}\bigr] \partial\!\!\!/ \bigl[ \lambda_- H + \tfrac{i}{2}(\gamma^{\rho\sigma}\lambda_-) F_{\rho\sigma}\bigr] \\
& + \tfrac{1}{4}\alpha'\bigl[ \overline{\lambda}_- H - \tfrac{i}{2}(\overline{\lambda}_- \gamma^{\mu\nu}) F_{\mu\nu}\bigr] \partial\!\!\!/ \bigl[ \lambda_+ H - \tfrac{i}{2}(\gamma^{\rho\sigma}\lambda_+) F_{\rho\sigma}\bigr] \ . 
\end{aligned} \quad (6.1)$$

Here $\alpha'$ is a real constant, and the subscripts $\pm$ are for the chiralities: $\lambda_\pm \equiv (1/2)(I\pm\gamma_5)\lambda$, while the $\widetilde{F}$ is the dual of $F$ defined by $\widetilde{F}_{\mu\nu} \equiv (1/2)\epsilon_{\mu\nu}{}^{\rho\sigma} F_{\rho\sigma}$. The first two terms in $\mathcal{L}_{\text{SDBI}}$ are the standard DBI terms with the notation $(F^4)_\mu{}^\mu \equiv F_\mu{}^\nu F_\nu{}^\rho F_\rho{}^\sigma F_\sigma{}^\mu$, while all the remaining terms are their supersymmetrizations [16] in an explicit manner.

In principle, $\mathcal{L}_{\text{SDBI}}$ is obtained by using tensor calculus for chiral multiplets. In superspace language [3], if we identify $H$ with $D$, the component lagrangian (6.1) is proportional to $\int d^4\theta\, (W^{\alpha+} W_{\alpha+})(W^{\beta-} W_{\beta-})$ [16]. Therefore all we need is the $D$-component of the product of the two superfields $(W^{\alpha+} W_{\alpha+})$ and $(W^{\beta-} W_{\beta-})$ of the opposite chiralities, and replace all the $D$'s by the $H$'s.

In our off-shell formulation, the lagrangian $\mathcal{L}_{\text{SDBI}}$ can be added to $\mathcal{L}_0$ without losing the invariance of the total action $I_3 \equiv I_0 + I_{\text{SDBI}}$. Even though $\mathcal{L}_{\text{SDBI}}$ is nothing but the conventional SDBI action [16] with all the $D$'s replaced by the $H$'s, the $C_{\mu\nu\rho}$-field equation is no longer trivial due to its propagation. In other words, our system provides the very first non-trivial interactions for the massive propagating tensor field $C_{\mu\nu\rho}$ in 4D.

## 7. Superspace Reformulation

We have so far dealt only with component formulation. It is the next natural step to re-formulate our results in superspace [3]. In this paper, we consider only global superspace without supergravity. Also, since all of our multiplets are off-shell, we have to impose superfield equations from outside, in order to recover the component results in section 3.

The most basic and crucial relationship is

$$[\nabla_A, \nabla_B\} \varphi = T_{AB}{}^C \nabla_C \varphi + m F_{AB} \quad . \tag{7.1}$$

Here we use the supercoordinates $(Z^A) \equiv (X^a, \theta^\alpha)$, with the indices $A \equiv (a,\alpha)$, $B \equiv (b,\beta)$, $\cdots$, where $a, b, \cdots = 0, 1, 2, 3$ (or $\alpha, \beta \cdots = 1, 2, 3, 4$) are for bosonic (or Majorana fermionic) coordinates. We use the (anti)symmetrization convention $M_{[AB)} \equiv M_{AB} - (-1)^{AB} M_{BA}$ in superspace. Our superfield strengths are $H_{ABCD}$, $G_{ABC}$, $F_{AB}$ in addition to the usual supertorsion $T_{AB}{}^C$ and vanishing supercurvature $R_{ABc}{}^d$. Note that we need an extra superfield strength $L_{ABC}$ with the same index structure as $G_{ABC}$. This superfield will turn out to be indispensable in order to satisfy certain Bianchi identities (BIs). As usual in superspace, we assign the dimension $d = 0$ to the potential superfield with purely bosonic indices, such as $C_{abc}$, $B_{ab}$ and $A_a$, while $d = 1/2$ to fermionic fundamental fields, such as $\lambda_\alpha$ and $\chi_\alpha$.

Our BIs are given by

$$\tfrac{1}{24}\nabla_{[A} H_{BCDE)} - \tfrac{1}{12} T_{[AB|}{}^F H_{F|CDE)} - \tfrac{1}{12} L_{[ABC} F_{DE)} \equiv 0 \quad , \tag{7.2a}$$

$$\tfrac{1}{6}\nabla_{[A} G_{BCD)} - \tfrac{1}{4} T_{[AB|}{}^E G_{E|CD)} - \tfrac{1}{6} L_{[ABC|} \nabla_{|D)}\varphi - m H_{ABCD} \equiv 0 \quad , \tag{7.2b}$$

$$\tfrac{1}{6}\nabla_{[A} L_{BCD)} - \tfrac{1}{4} T_{[AB|}{}^E L_{E|CD)} \equiv 0 \quad , \tag{7.2c}$$

$$\tfrac{1}{2}\nabla_{[A} F_{BC)} - \tfrac{1}{2} T_{[AB|}{}^D F_{D|C)} \equiv 0 \quad , \tag{7.2d}$$

$$\tfrac{1}{2}\nabla_{[A} T_{BC)}{}^D - \tfrac{1}{2} T_{[AB|}{}^E T_{E|C)}{}^D - \tfrac{1}{4} R_{[AB|d}{}^f (\mathcal{M}_f{}^e)_{|C)}{}^D \equiv 0 \quad . \tag{7.2e}$$

The new superfield $L_{ABC}$ appearing in the $H$- and $G$-BIs is an 'auxiliary' superfield strength with no physical dynamics. On the other hand, the $mH$-term in the $G$-BI is expected from the component results. In this paper, each of (7.2) is respectively called $(ABCDE)_H$, $(ABCD)_G$, $(ABCD)_L$, $(ABC)_F$ and $(ABC, D)_T$-BIs for convenience sake.

Our superspace constraints are summarized as

$$T_{\alpha\beta}{}^c = +2(\gamma^c)_{\alpha\beta} \quad , \quad T_{\alpha\beta}{}^\gamma = T_{\alpha b}{}^c = T_{ab}{}^c = T_{\alpha b}{}^\gamma = T_{ab}{}^\gamma = 0 \quad , \tag{7.3a}$$

$$G_{\alpha\beta c} = +2(\gamma_c)_{\alpha\beta} \quad , \quad G_{\alpha\beta\gamma} = 0 \quad , \quad R_{ABc}{}^d = 0 \quad , \tag{7.3b}$$

$$L_{\alpha\beta c} = +2(\gamma_c)_{\alpha\beta} \quad , \quad L_{\alpha\beta\gamma} = L_{\alpha bc} = L_{abc} = 0 \quad , \tag{7.3c}$$

$$G_{\alpha bc} = -(\gamma_{bc})_\alpha{}^\delta \chi_\delta \equiv -(\gamma_{bc}\chi)_\alpha \quad , \quad \nabla_\alpha \varphi = -\chi_\alpha \quad , \tag{7.3d}$$

$$H_{\alpha bcd} = -(\gamma_{bcd}\lambda)_\alpha \ , \qquad H_{\alpha\beta cd} = H_{\alpha\beta\gamma d} = H_{\alpha\beta\gamma\delta} = 0 \ , \tag{7.3e}$$

$$\nabla_\alpha \lambda_\beta = +\tfrac{1}{2}(\gamma^{cd})_{\alpha\beta} F_{cd} + \tfrac{1}{24}(\gamma^{abcd})_{\alpha\beta} H_{abcd} \ , \tag{7.3f}$$

$$\nabla_\alpha \chi^\beta = -(\gamma^c)_{\alpha\beta}\nabla_c \varphi - \tfrac{1}{6}(\gamma^{abc})_{\alpha\beta} G_{abc} \ . \tag{7.3g}$$

Corresponding to component transformation (3.4), there arises no explicit $m$-dependent terms in these constraints.

These constraints satisfy all the BIs at the dimensions $d \leq 1$. In particular, the $H$-BI starts at $d = -1/2$ as the $(\alpha\beta\gamma\delta\epsilon)_H$-BI. The first non-trivial role played by the $L$-superfield strength is seen at $d = 1/2$. In the $(\alpha\beta\gamma de)_H$-BI, we see that a term proportional to $T_{(\alpha\beta|}{}^f H_{f|\gamma)de}$ is cancelled by another term proportional to $L_{(\alpha\beta|[d} F_{e]|\gamma)}$, as desired. Similarly, in the $(\alpha\beta\gamma d)_G$-BI, the $L\nabla\varphi$-term is playing an important role. These $L$-dependent terms are also important at $d = 1$, because they cancel all the unwanted terms in the $H$- and $G$-BIs.

As usual in superspace, the BIs at $d = 3/2$ give the spinorial derivatives on the superfield strengths:

$$\nabla_\alpha H_{bcde} = +\tfrac{1}{6}(\gamma_{[bcd}\nabla_{e]}\lambda)_\alpha \ , \tag{7.4a}$$

$$\nabla_\alpha G_{bcd} = -\tfrac{1}{2}(\gamma_{[bc}\nabla_{d]}\chi)_\alpha - m(\gamma_{bcd}\lambda)_\alpha \ , \tag{7.4b}$$

$$\nabla_\alpha F_{bc} = +(\gamma_{[b}\nabla_{c]}\lambda)_\alpha \ . \tag{7.4c}$$

Since we deal with an off-shell formulation, all the fermionic superfield equations should be imposed from the outside of the system. Complying with the component lagrangian (3.1), we input the $\lambda$ and $\chi$-superfield equations

$$(\nabla\!\!\!\!/\,\lambda + m\chi)_\alpha \doteq 0 \ , \tag{7.5a}$$

$$(\nabla\!\!\!\!/\,\chi + m\lambda)_\alpha \doteq 0 \ . \tag{7.5b}$$

Taking spinorial derivatives of these equations, we get all the bosonic superfield equations:

$$(\gamma_{abc})^{\alpha\beta}\nabla_\alpha(\nabla\!\!\!\!/\,\lambda + m\chi)_\beta = +4(\nabla_d H_{abc}{}^d + mG_{abc}) \doteq 0 \ , \tag{7.6a}$$

$$(\gamma_a)^{\alpha\beta}\nabla_\alpha(\nabla\!\!\!\!/\,\lambda + m\chi)_\beta = +4(\nabla_b F_a{}^b + m\nabla_a\varphi) \doteq 0 \ , \tag{7.6b}$$

$$(\gamma_{ab})^{\alpha\beta}\nabla_\alpha(\nabla\!\!\!\!/\,\chi + m\lambda)_\beta = +4\nabla_c G_{ab}{}^c \doteq 0 \ . \tag{7.6c}$$

$$\nabla^\alpha(\nabla\!\!\!\!/\,\chi + m\lambda)_\alpha = -4\nabla_a^2 \varphi \doteq 0 \ , \tag{7.6d}$$

All of these superfield equations provide good supporting evidence for the consistency with the component results under supersymmetry. In particular, all the $m$-explicit terms are consistent with supersymmetry, despite its non-trivial features associated with compensators, including the two-form field $B_{ab}$ eventually absorbed into $C_{abc}$.

In conventional superspace, it has been a common wisdom that the auxiliary field $D$ for a VM arises out of the $\theta^4$-sector of a real scalar superfield $V$. Our result here with the 'auxiliary' field $C_{abc}$ provides a completely new viewpoint for a VM. Our result strongly indicates significant ingredients in superspace that have been overlooked for more than three decades since the first discovery of supersymmetry [20].

## 8. Concluding Remarks and Summary

In this paper, we have presented a new supersymmetric theory of a dilaton and a two-form field both of which play roles of compensators at the same time. The absorption of a dilaton $\varphi$ into a vector field $A_\mu$ by itself is not entirely new, because it is much like the usual compensator formalism [8][10][11]. However, the absorption of the two-form field $B_{\mu\nu}$ into a three-form field $C_{\mu\nu\rho}$, generating a mass for the latter is a new mechanism presented in this paper. The common wisdom keeps telling that a three-form field $C_{\mu\nu\rho}$ is to be a 'frozen' field in 4D without any propagating degrees of freedom. However, in our formulation, $B_{\mu\nu}$ is absorbed into the longitudinal components of $C_{\mu\nu\rho}$, making the latter propagate as a massive spinless field. The total consistency of our system is also guaranteed by global supersymmetry.

Notice not only that we have obtained the mass term for the three-form tensor $C_{\mu\nu\rho}$, but also that the mass term is non-tachyonic and physical. This gives other supporting evidence that our approach is on the right track for the mechanism for the two-form field as a compensator. We have also shown that our system can be further coupled to SDBI action, which gives a non-trivial confirmation of the physical significance of our system.

Note also that the field $C_{\mu\nu\rho}$ is originally 'auxiliary', but it starts propagating after the absorption of the compensator field $B_{\mu\nu}$. This phenomena is not entirely new, because in certain contexts of supersymmetric theories, auxiliary fields start propagating. Explicit examples are such as the multiplet of Lorentz connection, where some non-minimal auxiliary fields by Breitenlohner [21] start propagating [22], or in the theory of (curvature)$^2$-terms in supergravity in 3D, where even the graviton starts propagating, after (curvature)$^2$-terms are added [23]. Even though there are such analogs, the mechanism presented in this paper has also difference, because it deals with the massive propagating three-form field $C_{\mu\nu\rho}$ accompanied by $B_{\mu\nu}$ as a compensator.

By analyzing the $C_{\mu\nu\rho}$-field equation, we have found that our system maintains supersymmetry, even after adding a Fayet-Iliopoulos-like term $\mathcal{L}_{\xi H}$ [18]. This result is based on the peculiar feature that even though only the dual $H \equiv (1/4!)\epsilon^{\mu\nu\rho\sigma}H_{\mu\nu\rho\sigma}$ enters the lagrangian, the $C_{\mu\nu\rho}$-field equation has one derivative higher than the auxiliary $D$-field equation in the conventional system. We have seen that all the possible solutions $H \doteq H_0$, only $H \doteq 0$ is singled out for minimization of energy $T^{00}$, and supersymmetry is maintained.

We have also shown that our theory can have consistent interactions under supersymmetry, such as in the lagrangians $\mathcal{L}_{\rm CM}$ and $\mathcal{L}_{\rm SDBI}$. Subsequently, we have reformulated the results in section 3 in superspace. We have found the importance of the new 'auxiliary' superfield strength $L_{ABC}$ with no dynamics. Its non-vanishing component is $L_{\alpha\beta c}$ alone, with no physical degree of freedom. This $L$ is involved in highly non-trivial way, such as the $LF$-term in $H$-BIs, and also in the $L\nabla\varphi$-term in $G$-BIs.

Finally, we end with a brief summary of our work. There are six major new points in our formulation. First, the two-form field $B_{\mu\nu}$ plays a role of a compensator absorbed into the three-form tensor $C_{\mu\nu\rho}$, making the latter massive. Second, the usual pseudo-scalar auxiliary field $D$ can be replaced by its 'dual' three-form field $C_{\mu\nu\rho}$ which absorbs the $B_{\mu\nu}$ in the LM. Third, this mechanism works consistently with global supersymmetry. In particular, we discovered the new VM $(A_\mu, \lambda; C_{\mu\nu\rho})$. Fourth, our system works not only at the free-field level, but also with interactions, confirmed by the couplings to chiral multiplets and a SDBI action. Fifth, despite the parallel structure between the conventional auxiliary field $D$ and our $H \equiv (1/4!)\epsilon^{\mu\nu\rho\sigma}H_{\mu\nu\rho\sigma}$, there still are non-trivial differences due to the one higher derivative in the $C$-field equation. Sixth, in the superspace reformulation, we have discovered the new 'auxiliary' superfield strength $L_{ABC}$ with no dynamics. It is to be stressed that this peculiar role played by $L_{ABC}$ has not been presented in the past.